\documentclass{elsart}
\usepackage{graphicx}

\begin{document}
\begin{frontmatter}
\title  {Study of Possible Scintillation Mechanism Damage in
PbWO$_4$
           Crystals After Pion Irradiation}

\author[IHEP]{V.A.~Batarin},
\author[FNAL]{J.~Butler},
\author[NAN]{T.Y.~Chen},
\author[IHEP]{A.M.~Davidenko},
\author[IHEP]{A.A.~Derevschikov},
\author[IHEP]{Y.M.~Goncharenko},
\author[IHEP]{V.N.~Grishin},
\author[IHEP]{V.A.~Kachanov},
\author[IHEP]{A.S.~Konstantinov},
\author[IHEP]{V.I.~Kravtsov},
\author[IHEP]{V.A.~Kormilitsin},
\author[UMN]{Y.~Kubota},
\author[IHEP]{V.S.~Lukanin},
\author[IHEP]{Y.A.~Matulenko},
\author[IHEP]{Y.M.~Melnick},
\author[IHEP]{A.P.~Meschanin},
\author[IHEP]{N.E.~Mikhalin},
\author[IHEP]{N.G.~Minaev\thanksref{addr}},
\thanks[addr]{corresponding author, email: minaev@mx.ihep.su}
\author[IHEP]{V.V.~Mochalov},
\author[IHEP]{D.A.~Morozov},
\author[IHEP]{L.V.~Nogach},
\author[IHEP]{A.V.~Ryazantsev},
\author[IHEP]{P.A.~Semenov},
\author[IHEP]{V.K.~Semenov},
\author[IHEP]{K.E.~Shestermanov},
\author[IHEP]{L.F.~Soloviev},
\author[SYR]{S.~Stone},
\author[IHEP]{A.V.~Uzunian},
\author[IHEP]{A.N.~Vasiliev},
\author[IHEP]{A.E.~Yakutin},
\author[FNAL]{J.~Yarba}
\collab{BTeV electromagnetic calorimeter group}
\date{\today}

\address[IHEP]{Institute for High Energy Physics, Protvino, Russia}
\address[FNAL]{Fermilab, Batavia, IL 60510, U.S.A.}
\address[NAN]{Nanjing University, Nanjing, China}
\address[UMN]{University of Minnesota, Minneapolis, MN 55455, U.S.A.}
\address[SYR]{Syracuse University, Syracuse, NY 13244-1130, U.S.A.}

\begin{abstract} {
We employed two independent methods to study possible damage to
the scintillation mechanism in lead tungstate crystals due to
irradiation by a 34 GeV pion beam. First, 10 crystals were
irradiated simultaneously over 30 hours by a narrow beam, so that
only a small region of each crystal was affected. We studied the
effect of the irradiation on the light output non-uniformity. If a
localized degradation was observed, it would indicate damage to
the scintillation mechanism. Secondly, we detected light output
using two phototubes attached to sides of a crystal. Since these
phototubes detect scintillation light only from a small localized
region, the effect of transmission loss should be minimal. We did
not see any statistically significant evidence for scintillation
mechanism damage with either method. The effect is consistent with
zero, and the upper limit is 0.5$\%$ at 95$\%$ C.L.
}
\end{abstract}
\end{frontmatter}

\section{Introduction}
A high precision electromagnetic calorimeter (EMCAL) will add to
the exciting physics capabilities of BTeV \cite{BTeV} by detecting
photons and electrons with excellent energy resolution
(${\sigma}E/E\simeq1.8\%/\sqrt{E}$) \cite{nim1,nim2}. It will
consist of more than 10,000 lead tungstate (PbWO$_4$) crystals.
One of the most important challenges to maintaining the high
intrinsic resolution this system is to keep the absolute energy
calibration better than 0.2\%.

The system consists of crystals glued to photomultiplier tubes
(PMT) and monitored using a pulsed light source connected to the
crystals with a thin transparent fiber. There are several inherent
sources of instability in this system. First of all the gain of
the PMT's may change; this we monitor using red light which is
significantly less sensitive to changes in the crystal 
transmission than blue light \cite{red_light}. Secondly, 
the light outputs from
the crystals may also change. For example, the light output
decreases by 2\% for a temperature increase of 1$^0$~C; we will
maintain the temperature of the crystals constant to 0.1$^0$~C to
reduce this effect to a manageable level.

Another important reason for the light output of the crystals to
change is irradiation. Many BTeV crystals will be exposed to
radiation of less than 1 rad/hour, but some will receive 20
rad/hour. Even when the radiation dose rate is only a few
rad/hour, the crystals will suffer radiation damage, and as a
result, the light output will decrease \cite{nim3}. The light
output loss is believed to be due to the degradation in the
transmission of light inside the crystal, and not due to the
degradation of the light emission. We can monitor the transmission
loss (TL) by measuring it directly using light from a stable light
source. On the other hand, if there is light emission loss (EL),
this cannot be measured easily.

The BTeV crystals will be calibrated using electrons from the
data, whose momenta are measured in the tracking devices (in-situ
calibration). Collecting enough electrons in each crystal will
require only several hours of running. In fact, those crystals
that will be exposed to high radiation and will likely to suffer
most severe damage can be calibrated as often as every hour,
because the electron collection rate is highest in those crystals.
Since the in-situ calibration measures the light output directly,
it reflects the losses in the emission as well as in the
transmission. Thus we need only worry about calibrating crystals
using the light source for the period of data taking between the
in-situ electron calibrations.

In this paper we investigate the possible change in light
emissions in the crystals which would be a more serious change
than the loss of transmission as the transmission does recover in
time \cite{nim3}
This paper report results of two such studies.

The transmission loss is thought to be due to formation of color
centers, each of them is caused by an electron trapped in a
crystal defect. The trapping happens when the electron absorbs
radiation energy and jumps to higher energy states which exist
around the crystal defect. Even though valence electrons in the
PbWO$_4$ (PWO) crystal need to absorb a relatively large energy
corresponding to a short wavelength photon to be able to jump to
higher allowable energy levels, this excited electron needs less
energy to do so, and the corresponding photon wavelengths often
fall in the visible range. This causes the crystal to absorb
light. For the transmission loss to last, these trapped electrons
must be in metastable states. In PWO crystals, the relevant
metastable states have lifetime short enough at room temperature
so that when the radiation dose is reduced, the transmission will
recover gradually (hours to 10's of hours).

Emission may be reduced if light-emitting atom is changed under irradiation,
due to possible transformation of the nucleus of an atom in the nuclear
reaction.
One of the consequences is that such transformation, if happens, is likely
to be of permanent nature, which would lead to a cumulative effect.

Indirectly, we have already addressed this issue in our previous
study \cite{nim3}. We measured non-uniformity of the light output
along the crystal by exposing the crystals to a muon beam
perpendicular to the axes of the crystals.  Then we rotated the
crystal matrix so their axes were along the beam and irradiated
the crystals with a pion beam for 10 days. After the irradiation
we rotated the crystal matrix by 90$^0$ again and made another
scan with muon beam. The non-uniformity of the light output in the
front part of crystal (3-10 radiation lengths), where dose rate
from pions was varying within a factor of two, was about
0.5$\%$/cm. The non-uniformity did not change after an absorbed
dose up to 4 krad, which caused the signal loss of up to 30$\%$.
It was an indirect proof that we did not see scintillation
mechanism damage.

In previous studies no evidence of EL was found in
PWO crystals after their irradiation with gamma-source \cite{zhu}.
In this paper we present two independent methods of direct study
of possible scintillation mechanism damage.


Dr. R.Y. Zhu
proposed the main idea of the first method \cite{pc}. This method
relies on the light output difference between irradiated and
non-irradiated zones of a long crystal, after exposition to a
narrow pion beam traveling in the transverse direction relative to
the crystal length. In this method we irradiate a small section of
a crystal along its length using a pion beam, and measure the
light collection uniformity both before the irradiation and
immediately afterward. If one assumes that the loss is due only to
transmission, the uniformity should not be affected very much
since the light from anywhere in the crystal will travel in the
crystal at least once. Thus, it is affected evenly regardless of
where the light is emitted in the crystal. However, if there is
loss due to EL, it should only effect light from the irradiated
section. Therefore, if one observes localized loss in the area
where the crystal was irradiated, such loss can be attributed to
EL.

In the second approach, we mounted two PMT's on the sides of a
crystal and one at the end of the crystal. We presumed that if a
muon travels along the axis of the crystal, only the light emitted
near PMT's will reach the PMT's. The reason is, if the crystal
surface is optically flat, light from the other part of the
crystal has to travel a very long distance before it reaches one
of the PMT's. If this assumption is correct, then the detected
light loss can be attributed to the EL effect.

The studies were carried out in the IHEP test beam facility
\cite{nim3}.

\section{ Method I}

\subsection{Experimental setup.}
\begin{figure}[b]
\centering
\includegraphics[width=0.60\textwidth]{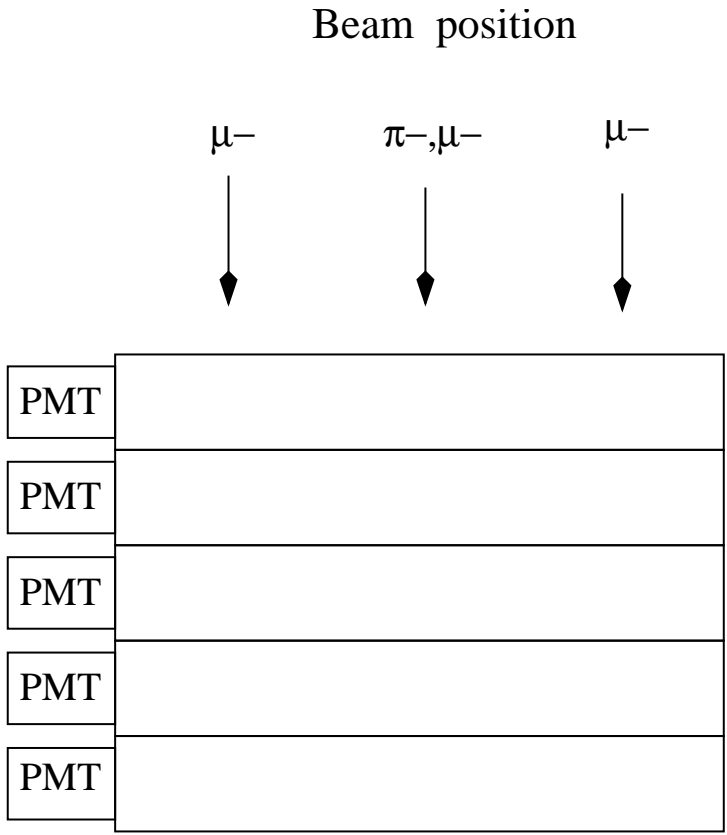}
\parbox{0.60\textwidth}{\caption{The layout of the crystals
in the first study and the beam positions (plan view). The crystals are 27 mm$^2$ in cross section and 220 mm long.
\label{fig:first_setup}}}
\end{figure}

In this study we used both pion and muon beams crossing an array
of PWO crystals in the perpendicular direction relative to the
crystal lengths. Fig.~\ref{fig:first_setup} shows how the crystals
were placed relative to the beam as seen from the top.
Five of the 10 crystals were produced by Shanghai Institute of Ceramics (SIC)
and the other five crystals were produced at Bogoroditsk Techno-Chemical
Plant (BTCP).
We used an intensive pion beam to irradiate only the middle section of the
crystals, while the muon beam was used to measure the light output
uniformity of the crystals along their lengths before and after irradiation.
To do this, the crystal array was moved so that the muon signal can be
measured in any area of a crystal.

The crystals were irradiated for about 30 hours with a beam intensity of
$2.6\cdot 10^6$ pions/spill. Each spill lasted for 1.5 sec, with the full
cycle of 9 seconds.
About $95\%$ of the pions were contained in an area of 2~cm in width
and 6~cm in height.

\subsection{Monte-Carlo simulations of dose rate}
\begin{figure}
\centering
\includegraphics[width=0.95\textwidth]{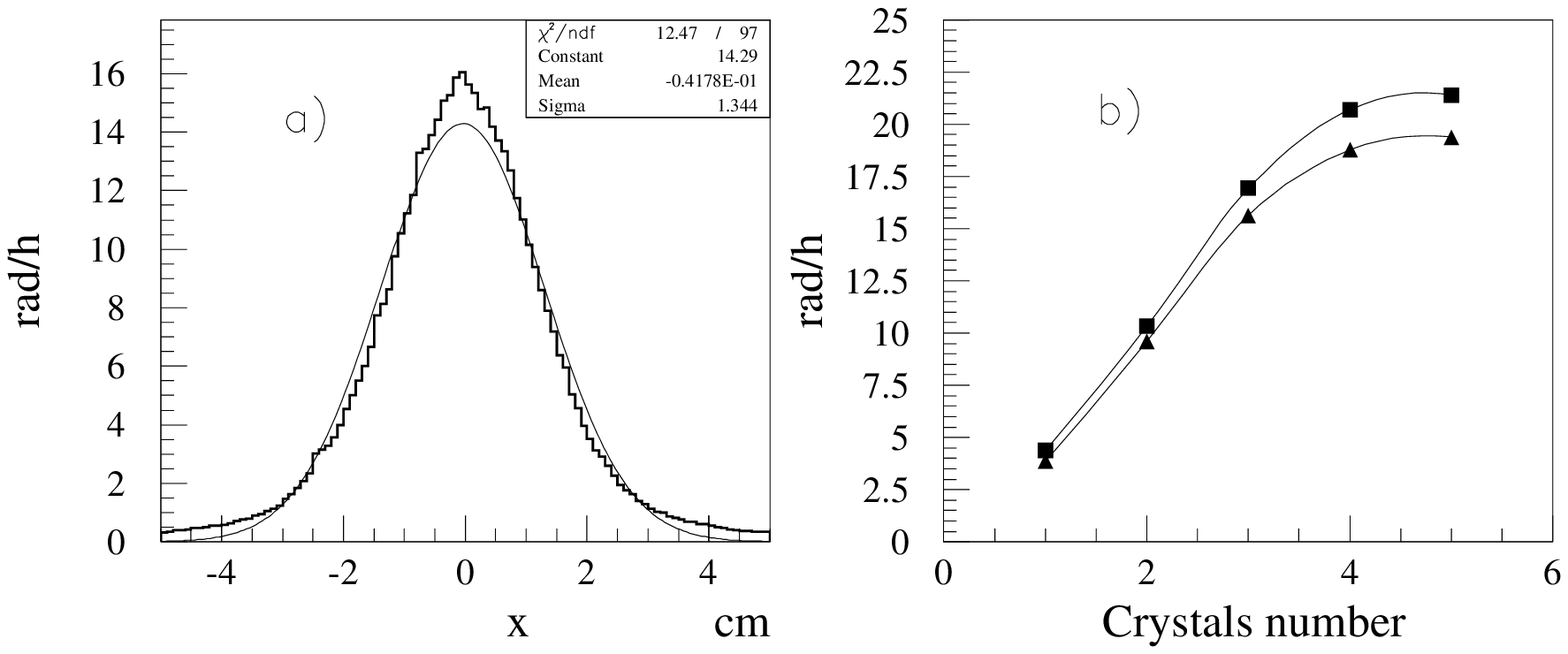}
\caption{a) The lateral dose rate  profile during the pion irradiation
runs for crystal 3 (BTCP).
The position $x=0$ corresponds to the center of the beam.
b) Maximum dose rate.
The filled squares indicate SIC crystals in the top layer, and filled
triangles indicate BTCP crystals.}
\label{fig:dose_profile}
\end{figure}

 To be able to predict how the light transparency inside the crystal varies,
we need to know the absorbed dose distributions of the pion beam.
The dose rate distributions along each crystal in the crystal matrix has
been studied by GEANT3 simulations.
Fig.~\ref{fig:dose_profile}a shows a lateral  dose rate profile during
the pion irradiation runs for crystal 3 (BTCP).
Also,  Fig.~\ref{fig:dose_profile}b shows the maximum dose rate for each
of the ten crystals.

\begin{figure}
\centering
\includegraphics[width=0.75\textwidth]{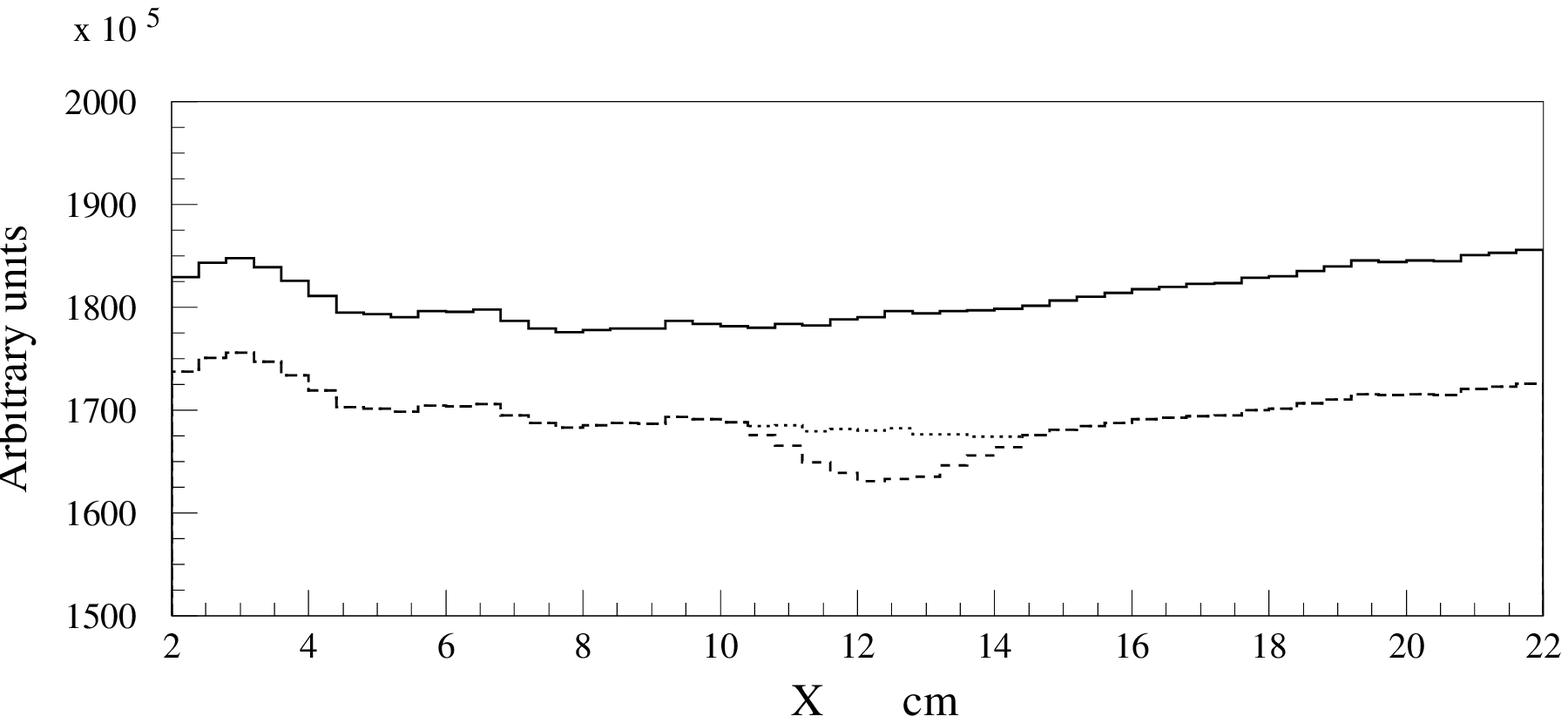}
\caption{Light response uniformity before (solid line) and after
(dashed line) irradiation. The dotted line indicates the response
when the scintillation mechanism damage was not included.
Monte-Carlo simulation.} \label{fig:Light_response}
\end{figure}


To understand possible effects qualitatively, a light collection
model was developed and incorporated in the GEANT3 simulation in
order to study possible changes in the light response uniformity
due to scintillation mechanism damage. We assumed that crystals
have ideal optical properties, thus anisotropy and diffuse
reflection were not included.

 The light response uniformity depends on both light transmittance and the
 local light emission. The zone of the color center
 formation was assumed the same as shown in Fig.~\ref{fig:dose_profile},
 a Gaussian distribution
 with $\sigma$  equal to 1.3 cm. Relative light transmittance was estimated
 as the ratio of the crystal response to muons after and before irradiation.
For illustrative purposes we assumed that the scintillation
mechanism degradation was 3$\%$ at the point of the maximum
absorbed dose. A proportionality  between absorbed energy and
degradation of the scintillation mechanism was introduced.

Fig. ~\ref{fig:Light_response} presents results of the simulation.
The upper curve (solid) shows the light output as a function of
the distance to the PMT and the lower curve (dashed) shows the
light output after irradiation that results in a loss of
transmission and with 3\% damage to the scintillation mechanism.
The dotted section is the response without damage to the
scintillation mechanism.

\begin{figure}
\centering
\includegraphics[width=0.7\textwidth]{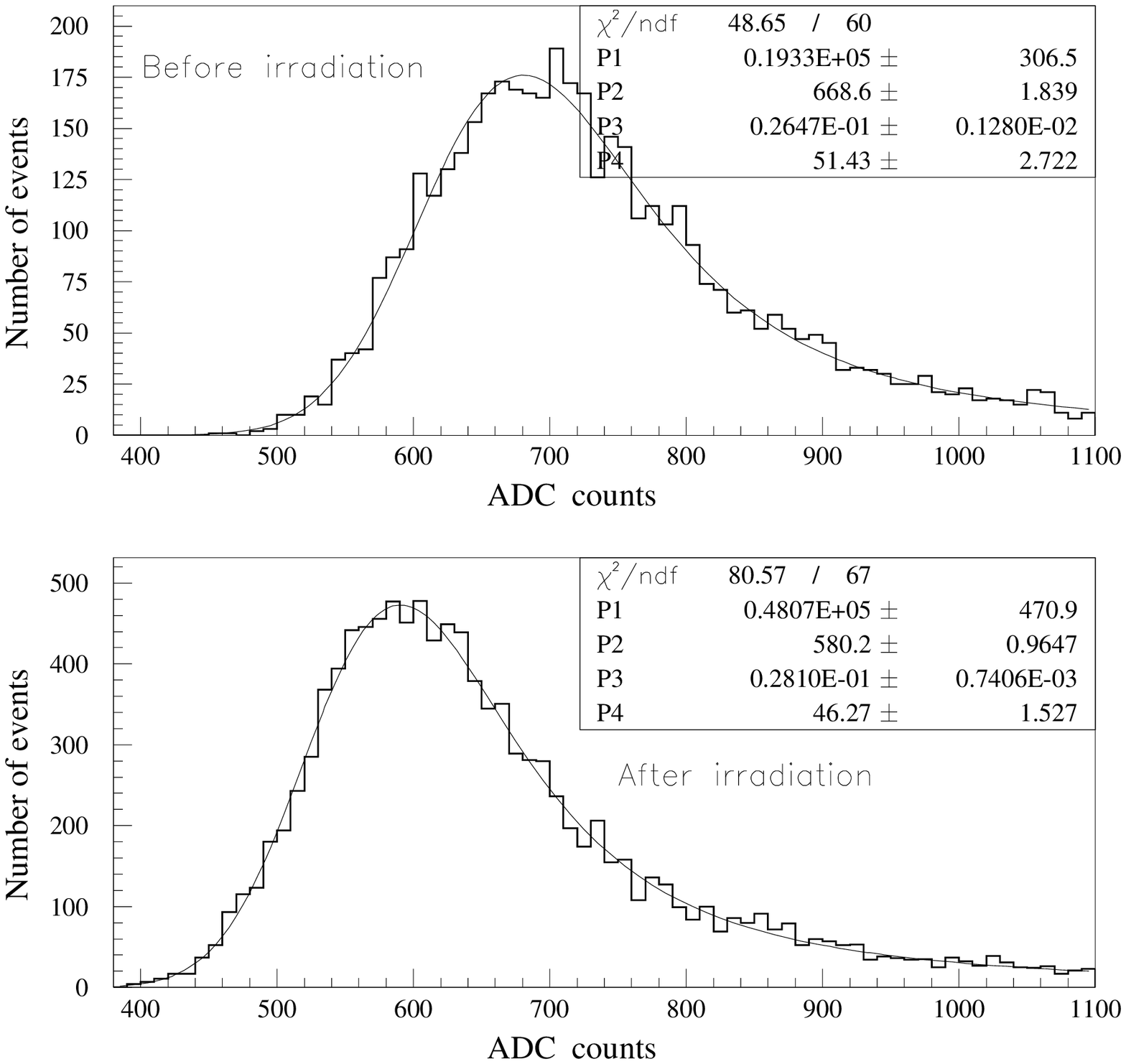}
\caption{Energy loss distributions for minimum ionizing particle
for crystal 2 (BTCP) before and after irradiation in the first
method. The data are fitted by a convolution of the Gaussian and
Landau distributions. The fit parameters are as follows : P1 - the
normalization factor, P2 - the most probable signal (MOP), P3 -
the inverse value of the Landau distribution FWHM, P4 - the
Gaussian sigma.} \label{fig:dau_gau}
\end{figure}

\subsection{ Results and discussion.}
The position of the muon track going through a crystal was
reconstructed using the drift chambers \cite{nim1}. The
pulse-height distribution collected for each selected region along
the crystal length was  fitted by a convolution of the Landau and
the Gaussian distributions to obtain a peak position. An example
of how we determine the peak position of the energy loss distribution for minimum ionizing particles is shown in Fig.~\ref{fig:dau_gau}.

\begin{figure}
\centering
\includegraphics[width=0.7\textwidth]{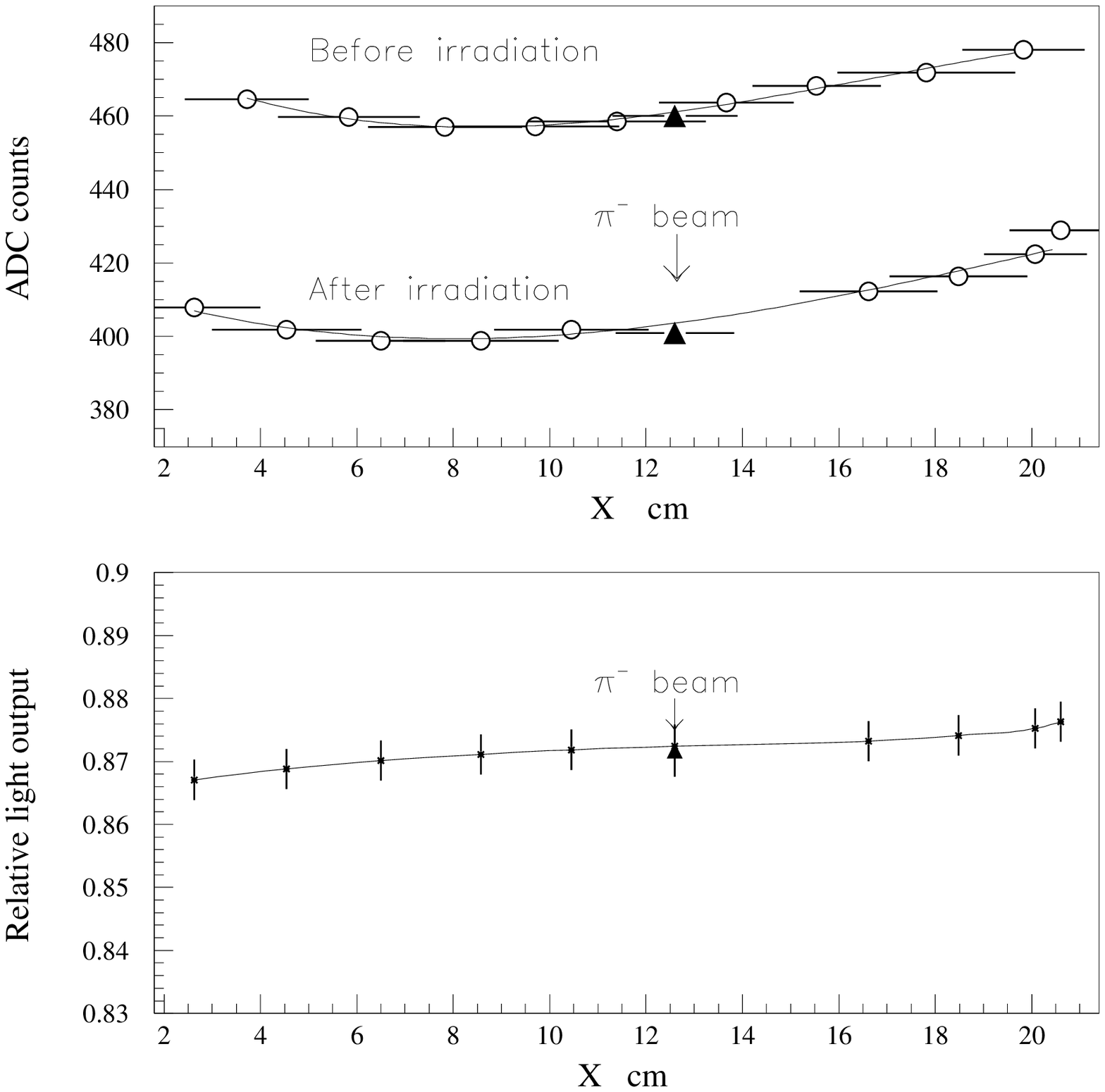}
\caption{ A typical  light response uniformity before and after irradiation.
 Maximun intensity for pion
 beam  is at the X=12.6 cm.   PMT position is at  X=0 cm.
}
\label{fig:unif_dist}
\end{figure}

After an estimation  of the muon peak position,  the light
response curves for different crystals were obtained. The results
for one such crystal from BTCP  are shown in
Fig.~\ref{fig:unif_dist}. The data were fitted to the 3-rd degree
polynomial function. In this figure, the arrow and the solid
triangles mark the region of the crystal, from 11 cm to 15 cm,
irradiated by pions. It was excluded from the fit of the data
after irradiation. We observed no an additional light loss in this
area showing that we are not sensitive to any scintillation
damage, at our level of radiation.

  We assume  that one can factorize the function of the light
output at any scanning muon position x in the crystal into two
parts. The first part depends on only TL of the light, which goes
from the point x to the phototube directly and indirectly by
reflecting from the crystal end. The second part, s(EL,x),
depends on only EL at the beam position x.

\begin{figure}
\begin{tabular}{lr}
\centering
\includegraphics[width=0.45\textwidth]{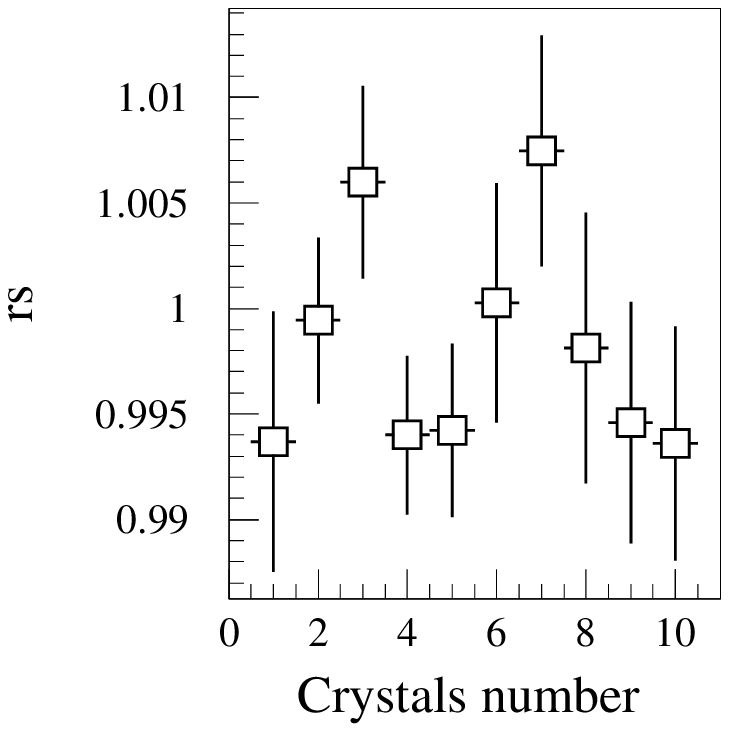}
&
\includegraphics[width=0.45\textwidth]{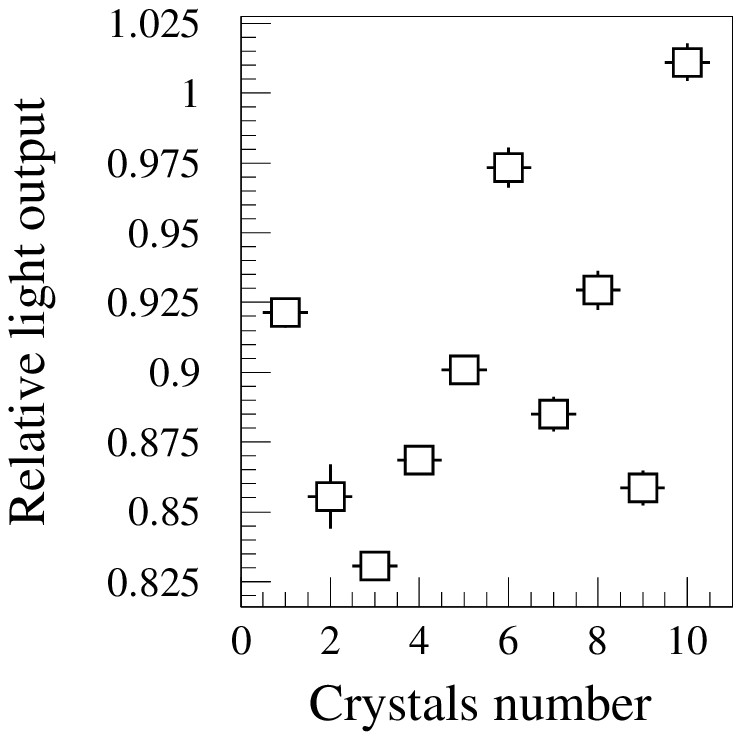}\\
\parbox{0.45\textwidth}{\caption{The relative emission light part of the total
 relative light output.
First five  crystals from Bogoroditsk, the second five crystals
from SIC (tapered). \label{fig:rellout}}}&
\parbox{0.45\textwidth}{\caption{The  total relative light output
in the zone of irradiation.
\label{fig:rlo}}}\\
\end{tabular}
\end{figure}

By taking the ratio of the light output as a function of x before
(index b) and after (index~a) radiation, we can isolate the effect
of loss of scintillation light.  The EL part of the relative light
output then is given by
\begin{equation}
rs=
\frac{s^a(EL,x_\pi)}{s^b(EL,x_\pi)},
\label{eq:r2_out}
\end{equation}
where $x_\pi$ is the pion beam position during the crystal
irradiation.

The $rs$ values for 10 crystals are shown in
Fig.~\ref{fig:rellout}. There is no light output alteration due to
scintillation mechanism damage  if $rs$ is equal to 1. In our
$\approx0.5\%$  accuracy range one can say, that for each out of
10 crystals we could not see an emission loss after pion
irradiation. The value  1 - $rs$ = (0.20$\pm$0.15)$\%$ when averaged over 
these 10 crystals.  

We conclude that if  the scintillation mechanism damage exists,
the degradation  of intrinsic  scintillation
light yield after pion irradiation is less
than 0.5$\%$ at a confidence level of 95$\%$. Fig.~\ref{fig:rlo} demonstrates
a total relative light output in the zone of irradiation.

\section{ Method II}

\subsection{ Experimental setup. }

\begin{figure}[b]
\centering
\includegraphics[width=0.4\textwidth, height=60mm]{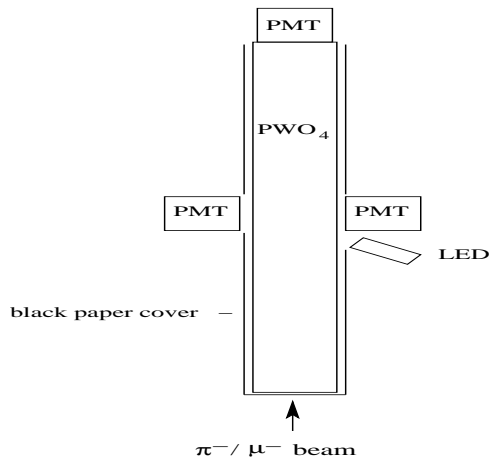} \\
\parbox{0.95\textwidth}{\caption{The layout of crystal and PMT's for the
direct-light measurements (plan view).
\label{fig:sam_setup}}}\\
\end{figure}

In this study, we attached 3 one-inch phototubes (Hamamatsu 5800)
to a crystal. Fig.~\ref{fig:sam_setup} shows the device layout.
The crystal was irradiated by a pion beam traveling along its
length. In addition, we used the same LED-based light transparency
monitoring system that we used at our test beam facility
\cite{nim4}. The crystal was wrapped in a black paper to minimize
diffusive reflections. The far PMT was coupled to the crystal
using optical grease, while the side PMTs were coupled with an air
gap. A light from the LED was injected into the crystal via an
optical fiber. The angle between the crystal's surface and the
optical fiber was 75$^{\circ}$. The side phototubes were placed
midway along the crystal.

Scintillation light produced near the side PMTs has to travel only
a few cm before it enters one of these PMTs. We assume that light
from other parts of the crystals won't even reach these PMTs.
Then, the effect of transparency degradation in the signal
detected by the side PMTs should be small. Transmission
degradation was monitored by the light source. We irradiated the
crystal for 8 hours with a beam intensity of $2.6\cdot 10^6$
pions/spill, which corresponds to the average dose rate in the
center of the crystal of $\approx15$rad/hour.

\subsection{ Results and discussion. }

\begin{figure}
\centering
\includegraphics[width=0.8\textwidth]{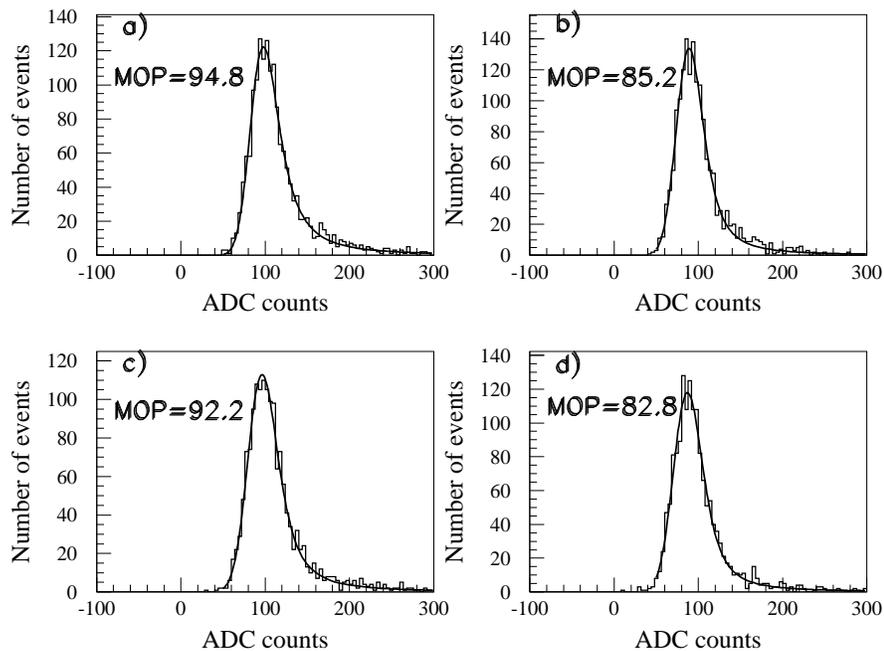}
\caption {Fit results for the distributions of the light outputs
which were detected by the side phototubes in the second method.
The same fit as in Fig.~\ref{fig:dau_gau} was used. a) Left PMT
before irradiation,
 c) left PMT after irradiation. b) right PMT before irradiation,
d) right PMT after irradiation. }
\label{fig:saml}
\end{figure}

In this method we compared crystal light output caused by muons
and measured by the side phototubes before and after irradiation.
This is similar to measurement of the light output from a thin
scintillating sample, when we do not sense light transmittance
degradation. A position of a muon track passing through the
crystal was reconstructed using the drift chambers. Muons in the
area of 4x4 mm$^2$ in the center of the crystal were used to
measure the light output.

 The  muon signal  distributions,
 fitted by the convolution of  Landau and  Gaussian  distributions, are
 shown in Fig.~\ref{fig:saml}.
The ratios of the amplitudes after irradiation to those before
irradiation for the muon and the LED signals are presented
in Table ~\ref{tab:ratio}. The LED signals did not change after
irradiation in either of the side phototubes.
This is expected, since we assume that the light going into the
side PMT's travel only small distances and the LED signal should be
insensitive to the transparency loss.
At the same time we did not find any statistically significant
light loss in the muon signals.
To treat the results of the second method correctly it's important to
take into account a contribution of the light from far points of the
muon track.

Because in the real life the crystal's surface is not perfectly flat,
there exists diffusive reflection of light in addition to the geometric
total internal reflection.
Some part of the scintillation light, particularly that which initially
travels in the direction parallel to the crystal length, may travel a long
 distance  before it reaches the side PMT's. This component will suffer
 transmission loss, even though the other component - light reaching the
 PMT's more directly - may not suffer transmission loss. We call
the former component, indirect light.
We can estimate its contribution into
the measured light output loss.

The third row of Table
~\ref{tab:ratio} shows attenuation of the LED signals in the far PMT,
although in the idealized case, light from the LED would not reach the far PMT.
 From the pulse-height spectra of LED signal, one can estimate the number of
photoelectrons if the variation is solely due to the photoelectron statistics.
Then the number of photoelectrons per ADC count equals the ratio
mean/$\sigma^2$.
 Table   ~\ref{tab:phel}
 presents portions of light in photoelectrons  from different sources.
 If we divide the number of photoelectrons for the far PMT by
the number of photoelectrons for the left PMT (see Table ~\ref{tab:phel}),
we obtain that
(0.7$\pm$0.1)$\%$ of blue LED light reaches the far PMT.
In our dedicated measurements at a stand after the accelerator run
we found that this mean value is about the same as the part of
indirect muon scintillation light which reaches the side PMTs.
We assign conservatively the error of our
estimate as 100\%. It gives us (0.7$\pm$0.7)$\%$ for this part.
From our previous measurements, we learned that the light input into a PMT
drops
by a factor of two if there is an air gap between
a crystal and a photocathode instead of using optical grease.
This decreases our estimated value of the contribution of indirect light in
the side PMT's from muon scintillation from (0.7$\pm$0.7)\% down to
(0.35$\pm$0.35)\%.

This suggests that the indirect light going into the side PMT's is
0.35\% of the light going into the far PMT.  {\em i.e.}
of about 50 photoelectrons in the side PMT's, (3.6$\pm$3.6)
(or (7.2$\pm$7.2)\%)  are indirect.
From the LED data, we estimate that the indirect light into the far PMT
lost 10.5\% after irradiation.
Assuming that the same fractional loss applies to the indirect light into
the side PMT's, we conclude that
(0.76$\pm$0.76)$\%$ of the light in the side PMT's will be lost due to the
transmission loss.
Subtracting this number from (2.8$\pm$1.4)$\%$ (see Table ~\ref{tab:ratio}),
we can finally estimate the light loss as (2.0$\pm$1.6)$\%$ which is
consistent with zero.

\begin{table}

\caption{The ratios of the amplitudes for muon and LED signals
after and before pion irradiation in the second method.}
\centering
\begin{tabular}{lccccc}  \hline
PMT   &    muon  & Blue LED \\\hline
Left  &0.973$\pm$0.013&1.000$\pm$0.001\\
Right &0.972$\pm$0.014&1.007$\pm$0.002\\
Far   &0.920$\pm$0.007&0.895$\pm$0.009\\\hline
\end{tabular}
\label{tab:ratio}
\end{table}

\begin{table}
\caption{Estimation of the numbers of photoelectrons before irradiation
in the second method.}
\centering
\begin{tabular}{lccc}  \hline
PMT   &    muon        & Blue LED  \\\hline
Left  &50.0$\pm$3.8 &7000$\pm$500\\
Right &50.9$\pm$3.2 &5200$\pm$300\\
Far   &1039$\pm$73    &45.6$\pm$2.9\\\hline
\end{tabular}
\label{tab:phel}
\end{table}

\section{Conclusions}

The first direct studies of possible damage to the scintillation mechanism
of lead tungstate crystals after hadron irradiation at
moderate dose rates was carried out for the BTeV experiment. We
have studied possible effects using two independent methods. We
did not see any evidence of the scintillation mechanism damage
over period of 8 to 30 hours in the either method. This effect is
consistent with zero and the upper limit for the absorbed dose up
to 600 rad is 0.5$\%$ at 95$\%$ confidence level.

This justifies the BTeV EMCAL calibration scenario that will rely
on the in-situ calibration with particles produced in physics
events, at least once a day, and on using monitoring with a light
pulser in between.


\section{Acknowledgements}

We thank the IHEP management for providing us infrastructure
support and accelerator time and to Fermilab for providing
equipment for data acquisition. Special thanks to Dr. R.Y. Zhu for
his valuable suggestions. This work was partially supported by the
U.S. National Science Foundation and the Department of Energy as
well as the Russian Foundation for Basic Research grant
02-02-39008.

\end{document}